\documentclass[fleqn,11pt]{ices05}
\usepackage{graphicx}

\begin{document}
{\small

\title{Development of a BEM Solver using Exact Expressions for Computing the
Influence of Singularity Distributions}

\author{S.Mukhopadhyay$^1$, N.Majumdar\footnote{Saha Institute of Nuclear
		Physics, 1/AF, Sector 1, Bidhannagar, Kolkata 700064, West Bengal,
		India, e-mail: supratik.mukhopadhyay@saha.ac.in}
	}
}

\abstract{Closed form expressions for three-dimensional potential and force
field due to singularities distributed on a finite flat surface have been used
to develop a fast and accurate BEM solver. The exact expressions have been
investigated in detail and found to be valid throughout the complete physical
domain. Thus, it has been possible to precisely simulate the complicated
characteristics of the potential and force field in the near-field domain.
The BEM solver  has been used to compute the capacitance of a unit square
plate and a unit cube. The obtained values have compared well with very
accurate values of the capacitances available in the literature.}

\section{Introduction}
The boundary element method is based on the numerical implementation of
boundary integral equations based on the Green's formula. In order to carry
out the implementation, the boundaries are generally segmented, and these
boundary elements are endowed with distribution of singularities such as
source, sink, dipoles and vortices. The singularity strengths are obtained
through the satisfaction of boundary conditions (Dirichlet, Neumann or Robin)
which calls for the computation of the influence of the singularities at the
points where the boundary conditions are being satisfied. Once the singularity
strengths are known, physical properties at any point within the physical
domain can be easily estimated. Thus, accurate computation of the influence
at a point in the domain of interest due to singularities distributed on a
surface is of crucial importance.

In this work, we have presented a BEM solver based on closed form expressions
of potential and force field due to a uniform distribution of source on a flat
surface. In order to validate the expressions, the potential and force field
in the near-field domain have been thoroughly investigated. In the process, the
sharp changes and discontinuities which characterize the near-field domain have
been easily reproduced.
Since the expressions are analytic and valid for the complete physical
domain, and no approximations regarding the size or shape of the singular
surface have been made during their derivation, their application is not
limited by the proximity of other singular surfaces or their curvature.
Moreover, since both potential and force fields are
evaluated using exact expressions, boundary conditions of any type, namely,
Dirichlet (potential), Neumann (gradient of potential) or Robin (mixed) can be
seamlessly handled.

As an application of the new solver, we have computed the capacitances of a
unit square plate and a unit cube to very high precision. These problems have
been considered to be two major unsolved problems of the electrostatic theory
\cite{Maxwell}-\cite{Mascagni}.
The capacitance values estimated by the present method have been compared with
very accurate results available in the literature (using BEM and other methods).
The comparison testifies to the accuracy of the approach and the developed
solver.

\section{Theory}
The expression for potential (V) at a point $(X, Y, Z)$ in free space due to
uniform
source distributed on a rectangular flat surface having corners situated at
$(x_1, z_1)$ and $(x_2, z_2)$ can be shown to be a multiple of
\begin{equation}
V(X,Y,Z) = \int_{z_1}^{z_2} \int_{x_1}^{x_2}
            \frac{dx\,dz}{\sqrt{(X-x)^2 + Y^2 + (Z-z)^2}}
\label{eq:Pot1}
\end{equation}
where the value of the multiple depends upon the strength of the source and
other physical considerations. Here, it has been assumed that the origin of
the coordinate system lies on the surface plane ($XZ$).
The closed form expression for $V(X,Y,Z)$ is as follows:
\begin{eqnarray}
\lefteqn{V(X,Y,Z) =} \nonumber \\
&& \frac{1}{2} \times \nonumber \\
&& \left( \right. + 2\,Z\,\ln \left( \frac{D_2\, -\, (X-x_1)} {D_1\, -\, (X-x_1)} \right) + 2\,Z\,\ln \left( \frac{D_3\, -\, (X-x_2)} {D_4\, -\, (X-x_2)} \right) \nonumber \\
&& + 2\,x_1\,\ln \left( \frac{D_1\, -\, (Z-z_1)} {D_2\, -\, (Z-z_2)} \right) + 2\,x_2\,\ln \left( \frac{D_4\, -\, (Z-z_2)} {D_3\, -\, (Z-z_1)} \right) \nonumber \\
&& + 2\,z_1\,\ln \left( \frac{D_1\, -\, (X-x_1)} {D_3\, -\, (X-x_2)} \right) + 2\,z_2\,\ln \left( \frac{D_4\, -\, (X-x_2)} {D_2\, -\, (X-x_1)} \right) \nonumber \\
&& -\, S_1\, \left( X\, +\, i\, \left| Y \right| \right)\, tanh^{-1} \left( \frac {R_1 - i\, I_1} {D_1\, \left| Z - z_1 \right|}  \right)
-\, S_1\, \left( X\, - i\, \left| Y \right| \right)\, tanh^{-1} \left( \frac {R_1 + i\, I_1} {D_1\, \left| Z - z_1 \right|}  \right) \nonumber \\
&& +\, S_2\, \left( X\, + i\, \left| Y \right| \right)\, tanh^{-1} \left( \frac {R_2 - i\, I_1} {D_2\, \left| Z - z_2 \right|}  \right)
+\, S_2\, \left( X\, - i\, \left| Y \right| \right)\, tanh^{-1} \left( \frac {R_2 + i\, I_1} {D_2\, \left| Z - z_2 \right|}  \right) \nonumber \\
&& +\, S_1\, \left( X\, + i\, \left| Y \right| \right)\, tanh^{-1} \left( \frac {R_1 - i\, I_2} {D_3\, \left| Z - z_1 \right|}  \right)
+\, S_1\, \left(  X\,  - i\, \left| Y \right| \right)\, tanh^{-1} \left( \frac {R_1 + i\, I_2} {D_3\, \left| Z - z_1 \right|}  \right) \nonumber \\
&& -\, S_2\, \left( X\,  + i\, \left| Y \right| \right)\, tanh^{-1} \left( \frac {R_2 - i\, I_2} {D_4\, \left| Z - z_2 \right|}  \right)
-\, S_2\, \left( X\,  - i\, \left| Y \right| \right)\, tanh^{-1} \left( \frac {R_2 + i\, I_2} {D_4\, \left| Z - z_2 \right|}  \right)\, \left. \right) \nonumber \\
&& - 2\,\pi\,Y
\end{eqnarray}
where
\begin{eqnarray*}
D_1 = \sqrt { (X-x_1)^2 + Y^2 + (Z-z_1)^2 };
D_2 = \sqrt { (X-x_1)^2 + Y^2 + (Z-z_2)^2 } \\
D_3 = \sqrt { (X-x_2)^2 + Y^2 + (Z-z_1)^2 };
D_4 = \sqrt { (X-x_2)^2 + Y^2 + (Z-z_2)^2 } \\
R_1 = Y^2 + (Z-z_1)^2;
R_2 = Y^2 + (Z-z_2)^2 \\
I_1 = (X-x_1)\,\left| Y \right|;
I_2 = (X-x_2)\,\left| Y \right|;
S_1 = {\it sign} (z_1-Z);
S_2 =  {\it sign} (z_2-Z)
\end{eqnarray*}

Similarly, the force components are multiples of
\begin{equation}
F_x(X,Y,Z) =
\ln \left( \frac{D_1\, - \, (Z-z_1)}{D_2\, -\, (Z-z_2)} \right) -\, \ln \left( \frac{D_3\, - \, (Z-z_1)}{D_4\, -\, (Z-z_2)} \right)
\end{equation}
\begin{eqnarray}
\lefteqn{F_y(X,Y,Z) =} \nonumber \\
&& -\frac{1}{2}\, \it{i}\, sign\, (Y) \times \nonumber \\
&& \left( \right. S_2\, tanh^{-1} \left( {\frac {R_2\, + \it{i}\, I_2}{D_4\, \left| Z-z_2 \right| }} \right) -\, S_2\, tanh^{-1} \left( {\frac {R_2\, - \it{i}\, I_2}{D_4\, \left| Z-z_2 \right| }} \right) \nonumber \\
&& +\, S_1\, tanh^{-1} \left( {\frac {R_1\, - \it{i}\, I_2}{D_3\, \left| Z-z_1 \right| }} \right) -\, S_1\, tanh^{-1} \left( {\frac {R_1\, + \it{i}\, I_2}{D_3\, \left| Z-z_1 \right| }} \right) \nonumber \\
&& +\, S_2\, tanh^{-1} \left( {\frac {R_2\, - \it{i}\, I_1}{D_2\, \left| Z-z_2 \right| }} \right) -\, S_2\, tanh^{-1} \left( {\frac {R_2\, + \it{i}\, I_1}{D_2\, \left| Z-z_2 \right| }} \right) \nonumber \\
&& +\, S_1\, tanh^{-1} \left( {\frac {R_1\, + \it{i}\, I_1}{D_1\, \left| Z-z_1 \right| }} \right)\, -\, S_1\, tanh^{-1} \left( {\frac {R_1\, - \it{i}\, I_1}{D_1\, \left| Z-z_1 \right| }} \right)\, \left. \right) \nonumber \\
&& +\, \it{C}
\end{eqnarray}
\begin{equation}
F_z(X,Y,Z) =
\ln \left( \frac{D_1\, -\, (X-x_1)}{D_2\, -\, (X-x_1)} \right) -\, \ln \left( \frac{D_3\, -\, (X-x_2)}{D_4\, -\, (X-x_2)} \right)
\end{equation}
In Eq.(4), $C$ is a constant of integration as follows:
\[\it{C} = \left\{
\begin{array}{l l}
  0 & \quad \mbox{if outside the extent of the flat surface}\\
  2\, \pi & \quad \mbox{if inside the extent of the surface and Y $>$ 0}\\
  -2\, \pi & \quad \mbox{if inside the extent of the surface and Y $<$ 0} \end{array} \right. \]
A BEM solver has been developed based on Eqs. (2)-(5) which has been used to
compute
the capacitances of a unit square plate and a unit cube. It is well known that
the charge densities near the edges of these bodies are much higher than those
far away from the edges. In order to minimize the errors arising out of the
assumption of uniform charge density on each panel, we have progressively
reduced the segment size close to the edges using a simple polynomial
expression as used in \cite{Read}.

\section{Results}
\begin{figure}
\vspace{-0.5in}
\begin{center}
\includegraphics[height=2in,width=4in]{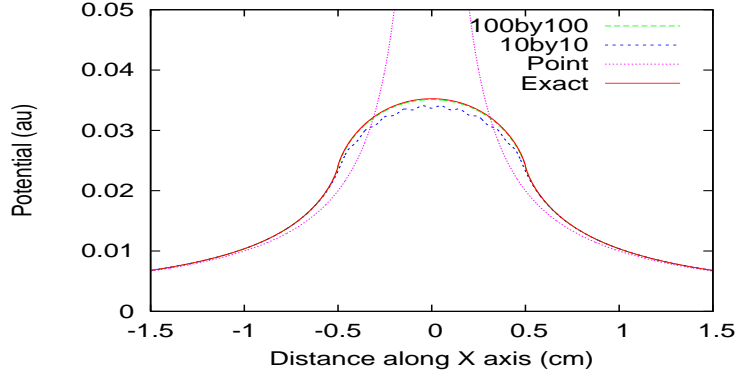}
\caption{\label{fig:potX} Comparison of potential distribution along X axis}
\end{center}
\end{figure}

\begin{figure}
\vspace{-0.5in}
\begin{center}
\includegraphics[height=2in,width=4in]{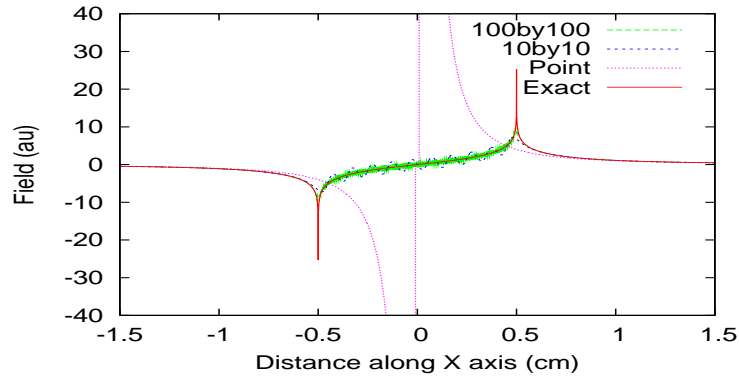}
\caption{\label{fig:EFX} Comparison of $E_x$ distribution along X axis}
\end{center}
\end{figure}

In order to establish the accuracy of the exact expressions, we have compared
the potential and electric field distributions computed using the new
expressions with those computed by assuming a varying degree of discretization
of the given surface, where each of the discrete element is assumed to have its
charge concentrated at its centroid. The flat surface has been assumed to be a
square ($1cm \times 1cm$) and length scale up to $10 \mu m$ has been resolved.
The strength of the source on the flat surface has been assumed to be unity.
In Figure\ref{fig:potX}, we have presented
a comparison among results obtained by the exact expressions and those obtained
by discretizing the flat surface having a single element, having 10$\times$10
elements and having 100$\times$100 elements. The maximum discretization
$100 \times 100$ discretization apparently yields good results even close to the
origin. However, oscillations in the potential value are visible with sufficient
close-up. Similar remarks are true for $E_y$, but for $E_x$
(Figure\ref{fig:EFX}), even the highest discretization leads to significant
amount of oscillation in the estimate. From these figures, we can also
conclude that the exact expressions reproduce the correct features of the
fields even in the most difficult situations.
The figures vividly represent the error incurred in modeling distributed sources
using the point analogy which is one of the most serious approximations of the
BEM.

\begin{figure}
\vspace{-0.5in}
\begin{center}
\includegraphics[height=4in,width=4in]{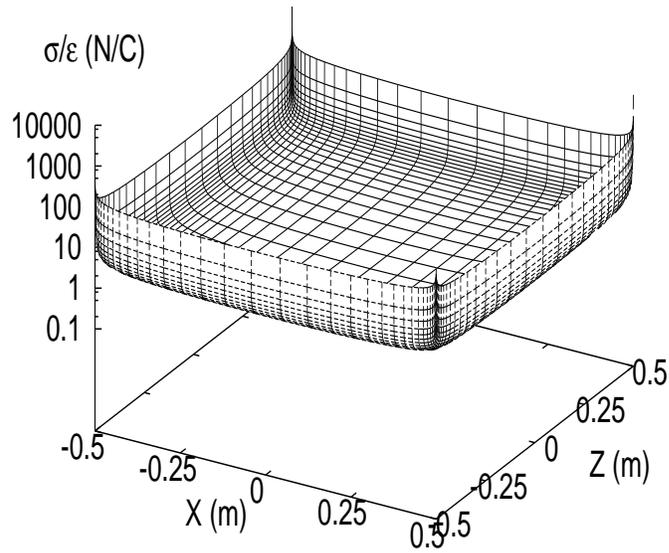}
\caption{\label{fig:Cube1} Variation of charge density on the top surface of the
cube}
\end{center}
\end{figure}

In Table \ref{table:table1}, we have presented a comparison of the values of
capacitances for a unit square plate and a unit cube as calculated by
\cite{Maxwell}-\cite{Mascagni}
and our estimations. Although it is difficult to comment regarding which is the
best result among the published ones, it is clear from the table that the
present solver indeed lead to very accurate results which are well within the
acceptable range.
\begin{table}
\vspace{-0.5in}
\centering
\begin{tabular}{| l | l | c | c |}
\hline
Reference & Method & Plate  & Cube \\
\hline
\cite{Maxwell} & Surface Charge & 0.3607 & - \\
\hline
\cite{Goto} & Refined Surface Charge & $0.3667892 \pm 1.1 \times 10^{-6}$ & $0.6606747 \pm 5 \times 10^{-7}$\\
& and Extrapolation & & \\
\hline
\cite{Read} & Refined Boundary Element & $0.3667874 \pm 1 \times 10^{-7}$ & $0.6606785 \pm 6 \times 10^{-7}$\\
& and Extrapolation & & \\
\hline
\cite{Mansfield} & Numerical Path Integration & 0.36684 & 0.66069 \\
\hline
\cite{Mascagni} & Random Walk on the Boundary & - & $0.6606780 \pm 2.7 \times 10^{-7}$ \\
\hline
This work & Boundary Element with & 0.3667869 & 0.6606746 \\
& Exact Expression for Potential & & \\
\hline
\end{tabular}
\caption{\label{table:table1}Comparison of capacitance values}
\end{table}
Finally, in figure \ref{fig:Cube1}, the variation of charge
density on the top surface of the unit cube has been presented. The sharp
increase in charge density near the edges and corners is quite apparent from
the figure.

\section{Conclusions}
A fast and accurate BEM solver has been developed using exact expressions for
potential and force field due to a uniform source distribution on a flat
surface. The expressions have been validated and found to yield very accurate
results in the complete physical domain. Of special importance is their ability
to reproduce the complicated field structure in the near-field region. The
errors incurred in assuming discrete point sources to represent a continuous
distribution have been illustrated. Accurate estimates of the capacitance of a
unit square plate and that of a unit cube have been made using the new solver.
Comparison of the obtained results with very accurate results available in the
literature has confirmed the accuracy of the approach and of the solver.

\reference{
\item \label{Maxwell}
J C Maxwell (1878):
\textit{Electrical Research of the Honorable Henry Cavendish}, 426 (Cambridge
University Press, Cambridge, 1878).

\item \label{Goto}
E Goto, Y Shi and N Yoshida (1992):
\textit{J Comput Phys}, 100, 105.

\item \label{Read}
F H Read (1997):
\textit{J Comput Phys}, 133, 1.

\item \label{Mansfield}
M L Mansfield, J F Douglas and E J Garboczi (2001):
\textit{Phys Rev E}, 64, 6, 61401.

\item \label{Mascagni}
M Mascagni and N A Simonov (2004):
\textit{J Comput Phys}, 195, 465.
}
\end{document}